\documentclass [11pt]{article}
\usepackage {amsmath,amssymb}

\makeatletter
\renewcommand\maketitle{%
  \begin{center}\Large\bf\@title\end{center}
  \begin{center}\@author\end{center}
  }
\makeatother

\newcounter{oftheorem}[section]
\newenvironment{mytheorem}[1]%
  {\begin{trivlist}
  \renewcommand{\theoftheorem}{#1~\thesection.\arabic{oftheorem}}
  \refstepcounter{oftheorem}
  \item[\hspace{\labelsep}\bf\thesection.\arabic{oftheorem}.\,#1.]}%
  {\end{trivlist}}

\newenvironment{definition}{\begin{mytheorem}{Definition}}{\end{mytheorem}}
\newenvironment{lemma}{\begin{mytheorem}{Lemma}\sl}{\end{mytheorem}}
\newenvironment{proposition}{\begin{mytheorem}{Proposition}\sl}{\end{mytheorem}}
\newenvironment{theorem}{\begin{mytheorem}{Theorem}\sl}{\end{mytheorem}}

\newenvironment{example}{\begin{mytheorem}{Example}}{\end{mytheorem}}

\newenvironment{proof}%
  {\begin{trivlist}
  \item[\hspace{\labelsep}\bf Proof.]}%
  {\end{trivlist}}

\newcounter{cislo}
\newenvironment{conditions}%
  {\begin{list}{\textup{(\arabic{cislo})}}{\usecounter{cislo}
   \itemsep0pt\topsep5pt\parsep\parskip
   \settowidth{\labelwidth}{\upshape(JP1)}\labelsep 1ex
   \leftmargin\labelwidth\addtolength{\leftmargin}{\labelsep}
   \addtolength{\leftmargin}{\parindent}}}%
  {\end{list}}

\def\0{{\bf 0}}                        
\def\1{{\bf 1}}                        
\def\N {\mathbb{N}}                    
\def\st {\,{:}\linebreak[0]\,\;}       
\def\ab {[\0,a] \cap [\0,b]}
\def\card {\mathop{{\rm card}}\nolimits}
\def\implies {\,$\Rightarrow$\,\linebreak[0]}

\allowdisplaybreaks[1]

\begin {document}

\title {Effect algebras with the maximality property}
\author {Josef Tkadlec\\[.5\baselineskip]
Department of Mathematics, Faculty of Electrical Engineering,\\
Czech Technical University, 166\,27 Praha, Czech Republic,\\
tkadlec@fel.cvut.cz}
\date {}

\maketitle

\begin {abstract}
The maximality property was introduced in~\cite {T:WS95} in orthomodular
posets as a common generalization of orthomodular lattices and orthocomplete
orthomodular posets. We show that various conditions used in the theory of
effect algebras are stronger than the maximality property, clear up the
connections between them and show some consequences of these conditions.
In particular, we prove that a Jauch--Piron effect algebra with a countable
unital set of states is an orthomodular lattice and that a unital set of
Jauch--Piron states on an effect algebra with the maximality property is
strongly order determining.
\end {abstract}

\section {Basic notions}

Effect algebras as generalizations of orthomodular posets (quantum logics)
are studied in the axiomatics of quantum systems---see, e.g.,~\cite {DP:NewTrends,
FB:Effect}.

\begin {definition}
An \emph {effect algebra} is an algebraic structure $(E,\oplus,\0,\1)$ such
that $E$ is a set, $\0$ and $\1$ are different elements of $E$ and $\oplus$
is a partial binary operation on $E$ such that for every $a,b,c \in E$ the
following conditions hold:
  \begin {conditions}
  \item $a \oplus b = b \oplus a$ if $a \oplus b$ exists,
  \item $(a \oplus b) \oplus c = a \oplus (b \oplus c)$ if $(a \oplus b)
        \oplus c$ exists,
  \item there is a unique $a'\in E$ such that $a \oplus a' = \1$ (\emph
        {orthosupplement}),
  \item $a=\0$ whenever $a \oplus \1$ is defined.
  \end {conditions}
\end {definition}

For simplicity, we use the notation $E$ for an effect algebra. A partial
ordering on an effect algebra $E$ is defined by $a \le b$ iff there is a $c
\in E$ such that $b = a \oplus c$. Such an element $c$ is unique (if it
exists) and is denoted by $b \ominus a$. $\0$ ($\1$, resp.) is the least
(the greatest, resp.) element of $E$ with respect to this partial ordering.
For every $a,b \in E$, $a''=a$ and $b' \le a'$ whenever $a \le b$. It can be
shown that $a \oplus \0 = a$ for every $a \in E$ and that a \emph
{cancellation law} is valid: for every $a,b,c \in E$ with $a \oplus b \le a
\oplus c$ we have $b \le c$. An \emph {orthogonality} relation on $E$ is
defined by $a \perp b$ iff $a \oplus b$ exists (iff $a \le b'$). (See,
e.g.,~\cite {DP:NewTrends,FB:Effect}.)

For $a \le b$ we denote $[a,b] = \{c \in E \st a \le c \le b \}$. A \emph
{chain} in $E$ is a nonempty linearly (totally) ordered subset of $E$.

Obviously, if $a \perp b$ and $a \lor b$ exist in an effect algebra, then
$a \lor b \le a \oplus b$. The reverse inequality need not be true (it holds
in orthomodular posets).

\begin {definition}
Let $E$ be an effect algebra. An element $a \in E$ is \emph {principal} if
$b \oplus c \le a$ for every $b,c \in E$ such that $b,c \le a$ and $b \perp
c$.
\end {definition}

\begin {definition}
An \emph {orthoalgebra} is an effect algebra $E$ in which, for every $a \in
E$, $a=\0$ whenever $a \oplus a$ is defined.

An \emph {orthomodular poset} is an effect algebra in which every element is
principal.

An \emph {orthomodular lattice} is an orthomodular poset that is a
lattice.
\end {definition}

Every orthomodular poset is an orthoalgebra. Indeed, if $a \oplus a$ is
defined then $a \oplus a \le a = a \oplus \0$ and, accordig to the
cancellation law, $a \le \0$ and therefore $a=\0$.

Orthoalgebras are characterized by the following conditions: the
orthosupplementation is an orthocomplementation (i.e., $a \lor a' = \1$ for
every $a$) or $a \oplus b$ is a minimal upper bound of $a,b$ for every
$a,b$. Orthomodular posets are characterized as effect algebras such that $a
\oplus b = a \lor b$ for every orthogonal pair $a,b$. (See~\cite {FB:Effect,
FGR:Filters}.) Let us remark that an orthomodular poset is usually defined
as a bounded partially ordered set with an orthocomplementation in which
the orthomodular law is valid.

Let us present a special class of orthomodular posets that we will use in
some examples.

\begin {proposition} \label {T:concreteOMP}
Let $X \neq \emptyset$, $E \subset \exp X$ be nonempty such that the
following conditions are fulfilled:
\begin {conditions}
\item $X \setminus A \in E$ whenever $A \in E$,
\item $A \cup B \in E$ whenever $A,B \in E$ are disjoint.
\end {conditions}
Then $(E,\oplus,\emptyset,X)$ with $A \oplus B = A \cup B$ for disjoint
$A,B \in E$ is an orthomodular poset such that the orthosupplement is the
set-theoretic complement and the partial ordering is the inclusion.
\end {proposition}

\begin {proof}
Since $E$ is nonempty, there is an element $A \in E$. According to the
condition~(1), $X \setminus A \in E$. According to the condition~(2), $X = A
\cup (X \setminus A) \in E$ and, according to the condition~(1), $\emptyset
= X \setminus X \in E$. It is easy to see that the axioms of an effect
algebra are fulfilled, that the orthosupplement is the set-theoretic
complement, that the partial ordering is the inclusion and that every
element of $E$ is principal.
\end {proof}

\begin {definition}
An orthomodular poset of the form of \ref {T:concreteOMP} is called \emph
{concrete}.
\end {definition}

Let us present two important notions we will use in the sequel.

\begin {definition}
A system $(a_i)_{i \in I}$ of (not necessarilly distinct) elements of an
effect algebra $E$ is \emph {orthogonal} if $\bigoplus_{i \in F} a_i$ is
defined for every finite set $F \subset I$.

An effect algebra $E$ is \emph {orthocomplete} if for every orthogonal
system $(a_i)_{i \in I}$ of elements of $E$ the supremum $\bigvee
\{\bigoplus_{i\in F} a_i \st F \subset I \ \text {is finite}\}$ exists.
\end {definition}

\begin {definition}
An effect algebra $E$ has the \emph {maximality property} if $\ab$ has a
maximal element for every $a,b \in E$.
\end {definition}

Obviously, every finite effect algebra has the maximality property and every
lattice effect algebra has the maximality property---$a \land b$ is a
maximal (even the greatest) element of $\ab$ for every $a,b$.

\section {States}

\begin {definition}
Let $E$ be an effect algebra. A \emph {state} $s$ on $E$ is a mapping $s
\st E \to [0,1]$ such that:
 \begin {conditions}
 \item $s(\1)=1$,
 \item $s(a \oplus b) = s(a) + s(b)$ whenever $a \oplus b$ is defined.
 \end {conditions}

A set $S$ of states on $E$ is \emph {unital}, if for every $a \in E
\setminus \{\0\}$ there is a state $s \in S$ such that $s(a) = 1$.

A set $S$ of states on $E$ is \emph {strongly order determining}, if for
every $a,b \in E$ with $a \not\le b$ there is a state $s \in S$ such that
$s(a) = 1 > s(b)$.
\end {definition}

Obviously, for every state $s$ we have $s(\0)=0$, $s(a')=1-s(a)$ for every
$a \in E$, $s(a) \le s(b)$ for every $a,b \in E$ with $a \le b$.

There are special two-valued states on concrete orthomodular posets (it is
easy to verify that they are indeed states):

\begin {definition}
Let $E \subset \exp X$ be a a concrete orthomodular poset, $x \in X$. The
state $s_x$ on $E$ defined by
  $$
  s_x(A) = \begin {cases}
           0 \,,& x \notin A \,,\\
           1 \,,& x \in A \,,
           \end {cases}
  \qquad A \in E \,,
  $$
is called \emph {carried by the point $x$}.
\end {definition}

It is easy to see that for concrete orthomodular posets the set of states
carried by points is strongly order determining.

\begin {lemma} \label {L:StronglyFull-Unital}
Every strongly order determining set of states on an effect algebra is
unital.
\end {lemma}

\begin {proof}
Let $S$ be a strongly order determining set of states on an effect algebra
$E$, $a$ be a nonzero element of $E$. Then $a \not\le \0$ and therefore
there is a state $s \in S$ such that $s(a) = 1 > s(\0)$.
\end {proof}

Let us present two observations describing the impact of a sufficiently
large state spaces to the properties of the algebraic structure.

\begin {proposition} \label {T:unital-OA}
Every effect algebra with a unital set of states is an orthoalgebra.
\end {proposition}

\begin {proof}
Let $E$ be an effect algebra with a unital set $S$ of states. Let $a \in E$
be such that $a \oplus a$ is defined. Then $1 \ge s (a \oplus a) = 2 \,
s(a)$ and therefore $s(a) < \frac 12$ for every state $s \in S$. Since $S$
is unital, we obtain that $a=\0$.
\end {proof}

\begin {proposition} \label {T:SOD->OMP}
Every effect algebra with a strongly order determining set of states is an
orthomodular poset.
\end {proposition}

\begin {proof}
Let $E$ be an effect algebra with a strongly order determining set $S$ of
states. Let us prove that every element of $E$ is principal. Let $a,b,c \in
E$ such that $b,c \le a$ and $b \perp c$. Then for every state $s \in S$
with $s(a')=1$ we consecutively obtain: $0 = s(a) = s(b) = s(c) = s(b\oplus
c)$, $s\bigl((b \oplus c)'\bigr) = 1$. Since the set $S$ is strongly order
determining, we obtain that $a' \le (b \oplus c)'$ and therefore $b \oplus c
\le a$.
\end {proof}

\section {Jauch--Pironness}

\begin {definition}
Let $E$ be an effect algebra. A state $s$ on $E$ is \emph {Jauch--Piron} if
for every $a,b \in E$ with $s(a)=s(b)=1$ there is a $c \in E$ such that $c
\le a,b$ and $s(c)=1$.

An effect algebra is \emph {Jauch--Piron} if every state on it is
Jauch--Piron.
\end {definition}

The following statement was proved in~\cite [Proposition 2.6]{T:CEOEA},
we will generalize it later (\ref {T:JPCU->OML}).

\begin {lemma} \label {T:JPCU->M}
Every Jauch--Piron effect algebra with a countable unital set of states
has the maximality property.
\end {lemma}

\begin {proof}
Let $E$ be a Jauch--Piron effect algebra with a countable unital set $S$ of
states. Let $a,b \in E$. If $\ab = \{\0\}$ then $\0$ is a maximal element of
$\ab$. Let us suppose that $\ab \neq \{\0\}$. Then there is an element $c
\in \ab \setminus \{\0\}$ and, since the set $S$ is unital, there is a state
$s \in S$ such that $s(c)=1$. Hence $s(a)=s(b)=1$ and the set $S_{a,b} = \{
s \in S \st s(a)=s(b)=1 \}$ is nonempty and countable. Let $s_0$ be a
$\sigma$-convex combination (with nonzero coefficients) of all states from
$S_{a,b}$. Then $s_0(a) = s_0(b) = 1$. Since the state $s_0$ is
Jauch--Piron, there is an element $c \in \ab$ such that $s_0(c)=1$. It
remains to prove that $c$ is a maximal element of $\ab$. Indeed, if $d \in
\ab$ with $d \ge c$ then $e = d \ominus c \in \ab$ and $e \perp c$. Hence
$s_0(e) = 0$ and therefore there is no state $s\in S$ such that $s(e)=1$.
Due to the unitality of $S$, $e=0$ and therefore $d=c$.
\end {proof}

\begin {proposition} \label {T:UJP->OMP}
Every effect algebra with the maximality property and with a unital set of
Jauch--Piron states is an orthomodular poset.
\end {proposition}

\begin {proof}
Let $E$ be an effect algebra with the maximality property and with a unital
set $S$ of Jauch--Piron states. Let us suppose that $E$ is not an
orthomodular poset and seek a contradiction. There are elements
$a,b,c \in E$ such that $b, c \le a$, $b \perp c$ and $b \oplus c \not\le
a$. Let us denote $d = b \oplus c$. Since $E$ has the maximality
property, there is a maximal element $e$ in $[\0,a'] \cap [\0,d']$. Since $d
\not\le a$, we obtain that $a' \not\le d'$ and therefore $e < a'$
and $a' \ominus e \neq \0$. Since the set $S$ is unital, there is a state $s
\in S$ such that $s(a' \ominus e) = 1$. Hence $s(a')=1$, $0 = s(e) = s(a) =
s(b) = s(c) = s(d)$, $s(d') = 1$, $s(d'\ominus e) = 1$. Since the state $s$ is
Jauch--Piron,
there is an element $f \in E$ such that $f \le (a' \ominus e), (d' \ominus
e)$ and $s(f) = 1$. Hence $f \neq \0$ and $e < e \oplus f \le a',d'$---this
contradicts to the maximality of $e$.
\end {proof}

Let us remark that there are effect algebras with the maximality property
that are not orthoalgebras---e.g., the 3-chain $C_3 = \{\0,a,\1\}$ with $a
\oplus a = \1$ and $x \oplus \0 = x$ for every $x \in C_3$. It seems to be
an open question whether the assumption of the maximality property in \ref
{T:UJP->OMP} might be omitted (it is not a consequence of the existence of
a countable unital set of Jauch--Piron states---see \ref {E:OMPnotM}.)
\ref {T:UJP->OMP} cannot be improved to orthomodular lattices---see \ref
{E:UnotSOD} ($\{\frac12\,(s_x+s_y)\st x,y \in X, \ x \neq y\}$ is a unital
set of Jauch--Piron states).

It is well-known and easy to see that every state on a Boolean algebra is
Jauch--Piron and that a unital set of states on a Boolean algebra is
strongly order determining. Let us generalize the latter statement.

\begin {theorem} \label {T:Unital-StronglyFull}
A set of Jauch--Piron states on an effect algebra with the maximality
property is unital if and only if it is strongly order determining.
\end {theorem}

\begin {proof}
$\Leftarrow$: See \ref {L:StronglyFull-Unital}.

$\Rightarrow$: Let $E$ be an effect algebra with the maximality property and
with a unital set $S$ of Jauch--Piron states. Let $a,b \in E$ such that $a
\not\le b$. Let $c \in E$ be a maximal element of $\ab$. Then $c<a$ and
therefore $a \ominus c \neq \0$. Since the set $S$ is unital, there is a
state $s \in S$ such that $s(a \ominus c) = 1$ and therefore $s(a)=1$. Let
us suppose that $s(b)=1$ and seek a contradiction. Since $s$ Jauch--Piron,
there is an element $d \in E$ such that $d \le a \ominus c$, $d \le b$ and
$s(d)=1$. Hence $d \neq \0$ and $c < c \oplus d \le a$.  According to \ref
{T:UJP->OMP}, $b$ is principal and therefore $c \oplus d \le b$---this
contradicts to the maximality of $c$.
\end {proof}

Let us remark that \ref {T:UJP->OMP} is a consequence of \ref
{T:Unital-StronglyFull} and \ref {T:SOD->OMP}. Let us present examples that
the assumptions in \ref {T:Unital-StronglyFull} cannot be omitted.

\begin {example} \label {E:UnotSOD}
Let $X=\{a,b,c,d\}$, $E$ be the family of even-element subsets of $X$ with
the $\oplus$ operation defined as the union of disjoint sets. Then
$(E,\oplus,\emptyset,X)$ is a finite (hence with the maximality property)
concrete orthomodular poset and the set $S = \{s_a, s_b, s_c\}$ of states
carried by points $a,b,c$ is a unital set of (two-valued) states on $E$ that
is not strongly order determining: $\{a,d\} \not\le \{a,b\}$ but there is no
state $s \in S$ such that $s(\{a,d\}) = 1 > s(\{a,b\})$. (States in $S$ are
not Jauch--Piron.)
\end {example}

\begin {example} \label {E:OMP-UnotSOD}
Let $X_1,X_2,X_3,X_4$ be nonempty mutually disjoint sets, $X_1, X_3$ be
infinite, $X = \bigcup_{i=1}^4 X_i$,
  \begin {align*}
  E_0 &= \{\emptyset, X_1 \cup X_2, X_2 \cup X_3, X_3 \cup X_4, X_4 \cup X_1, X\}\,,\\
  E   &= \{(A \setminus F) \cup (F \setminus A) \st F \subset X_1 \cup X_3 \
           \text {is finite},\ A \in E_0 \}\,,
  \end {align*}
$A \oplus B = A \cup B$ for disjoint $A,B \in E$. Then
$(E,\oplus,\emptyset,X)$ is a concrete orthomodular poset and the set
$S=\{s_x \st x \in X_1 \cup X_3\}$ of states carried by points from $X_1
\cup X_3$ is a unital set of (two-valued) Jauch--Piron states on $E$. The set
$S$ is not strongly order determining because $X_1 \cup X_4 \not\le X_1 \cup
X_2$ and for every $s \in S$ with $s(X_1 \cup X_4) = 1$ there is an $x \in
X_1$ such that $s = s_x$ and therefore $s(X_1 \cup X_2) = 1$. ($E$ does not
have the maximality property.)
\end {example}

\begin {theorem} \label {T:JPCU->OML}
Every Jauch--Piron effect algebra with a countable unital set of states is
an orthomodular lattice.
\end {theorem}

\begin {proof}
Let $E$ be a Jauch--Piron effect algebra with a countable unital set $S$ of
states. According to \ref {T:JPCU->M}, $E$ has the maximality property.
According to \ref {T:Unital-StronglyFull}, the set $S$ is strongly order
determining. According to \ref {T:SOD->OMP}, $E$ is an orthomodular poset.
Let us show that $a \land b$ exists for every $a,b \in E$. (Then also $a
\lor b = (a' \land b')'$ exists for every $a,b \in E$.) If $\ab=\{\0\}$ then
$\0 = a \land b$. Let us suppose that there is a nonzero element $c \in E$
such that $c \le a,b$. Then there is a state $s \in S$ such that $s(c)=1$.
Hence $s(a)=s(b)=1$ and the set $S_{a,b} = \{ s \in S \st s(a)=s(b)=1 \}$ is
nonempty and countable. Let $s_0$ be a $\sigma$-convex combination (with
nonzero coefficients) of all states from $S_{a,b}$. Then $s_0(a)=s_0(b)=1$.
Since the state $s_0$ is Jauch--Piron, there is an element $c_0 \in E$ such
that $c_0 \le a,b$ and $s_0(c_0)=1$. Hence $s(c_0)=1$ for every $s \in
S_{a,b}$. For every $c \in \ab$ and every $s \in S$ with $s(c)=1$ we have $s
\in S_{a,b}$ and therefore $s(c_0)=1$. Since $S$ is strongly order
determining, $c \le c_0$ for every $c \in \ab$. Hence $c_0 = a \land b$.
\end {proof}

Let us present examples that the conditions in \ref {T:JPCU->OML} cannot be
omitted. There is a concrete (hence with a strongly order determining set of
two-valued states) Jauch--Piron orthomodular poset that is not a
lattice---see~\cite {Muller} (every unital set of states on it is
uncountable). As the following example shows there is an orthomodular poset
with a countable strongly order determining set of (two-valued) Jauch--Piron
states that does not have the maximality property and therefore it is not a
lattice (there are non-Jauch--Piron states).

\begin {example} \label {E:OMPnotM}
Let $X_1,X_2,X_3,X_4$ be mutually disjoint countable infinite sets,
$X = \bigcup_{i=1}^4 X_i$,
  \begin {align*}
  E_0 &= \{\emptyset, X_1 \cup X_2, X_2 \cup X_3, X_3 \cup X_4, X_4 \cup X_1, X\}\,,\\
  E   &= \{(A \setminus F) \cup (F \setminus A) \st F \subset X \ \text {is
            finite},\ A \in E_0 \}\,,
  \end {align*}
$A \oplus B = A \cup B$ for disjoint $A,B \in E$. Then $(E, \oplus,
\emptyset, X)$ is a concrete orthomodular poset and the set $S=\{s_x \st x
\in X\}$ of states carried by points is a countable strongly order
determining set of two-valued Jauch--Piron states on $E$. The set
$[\emptyset, X_1 \cup X_2] \cap [\emptyset, X_4 \cup X_1]$ consists of
finite subsets of $X_1$, hence $E$ does not have the maximality property. As
an example of a non-Jauch--Piron state we can take a $\sigma$-convex
combination (with nonzero coefficients) of states from $\{s_x \st x \in
X_1\}$.
\end {example}

\section {Relationship of various conditions}

\begin {theorem} \label {T:MaximalityProperties}
Let $E$ be an effect algebra. Consider the following poperties:
  \begin {conditions}
  \item [\rm(F)] $E$ is finite.
  \item [\rm(CF)] $E$ is chain finite.
  \item [\rm(OC)] $E$ is orthocomplete.
  \item [{\makebox[0pt][r]{\rm(JPCU)}}] $E$ is Jauch--Piron with a
        countable unital set of states.
  \item [\rm(L)] $E$ is a lattice.
  \item [\rm(CU)] For every $a,b \in E$, every chain in $\ab$ has an upper
        bound in $\ab$.
  \item [\rm(M)] $E$ has the maximality property.
  \end {conditions}
Then the following implications hold:
\rm
(F)\implies(CF)\implies(OC)\implies(CU)\implies(M),
(JPCU)\implies(L)\implies(CU).
\end {theorem}

\begin {proof}
(F)\implies(CF): Obvious.

(CF)\implies(OC): Every orthogonal system in a chain finite effect algebra
is finite. Hence $E$ is orthocomplete.

(OC)\implies(CU): Let $C$ be a chain in $\ab$. According to~\cite [Theorem
3.2]{JP:Orthocomplete}, every chain in an orthocomplete effect algebra has a
supremum. This supremum obviously belongs to $\ab$.

(CU)\implies(M): Let $a,b \in E$. Since $\ab \supset \{\0\}$, the family of
chains in $\ab$ is nonempty. According to Zorn's lemma, there is a maximal
chain $C$ in $\ab$. According to the assumption, there is an upper bound $c
\in \ab$ of $C$. Since the chain $C$ is maximal, $c \in C$ is a maximal
element of $\ab$.

(JPCU)\implies(L): See \ref {T:JPCU->OML}.

(L)\implies(CU): Let $a,b \in E$. The element $a \land b$ is an upper bound
for every chain in $\ab$.

\end {proof}

Let us present examples that the scheme of implications in the previous
theorem cannot be improved.

\begin {example}
Let $X$ be an infinite set, $y \notin X$, $E = \{\emptyset\} \cup \bigl\{
\{x,y\} \st x \in X \bigr\} \cup \bigl\{ X \setminus \{x\} \st x \in X\}
\cup \bigl\{ X \cup \{y\} \bigr\}$, $A \oplus B = A \cup B$ for disjoint
$A,B \in E$. Then $(E,\oplus,\emptyset,X\cup\{y\})$ is an infinite chain
finite concrete orthomodular lattice.
\end {example}

\begin {example}
Let $X$ be an uncountable set, $E=\exp X$ with $A \oplus B = A \cup B$ for
disjoint $A,B \in E$. Then $(E,\oplus,\emptyset,X)$ is an orthocomplete
concrete orthomodular lattice (it forms a Boolean algebra) such that there
is an uncountable set of mutually orthogonal elements. Hence it is not chain
finite and every unital set of states on $E$ is uncountable.
\end {example}

\begin {example}
Let $X$ be a countable infinite set. Let $E$ be a family of finite and
cofinite subsets of $X$ with the $\oplus$ operation defined as the union of
disjoint sets. Then $(E,\oplus,\emptyset,X)$ is a concrete orthomodular
lattice (it forms a Boolean algebra) fulfilling the condition~(JPCU) (every
state on a Boolean algebra is Jauch--Piron, there is a countable unital set
of states carried by points) that is not orthocomplete.
\end {example}

\begin {example}
Let $X$ be a 6-element set. Let $E$ be the family of even-element subsets of
$X$ with the $\oplus$ operation defined as the union of disjoint sets from
$E$. Then $(E,\oplus,\emptyset,X)$ is a finite concrete orthomodular poset
that is not a latice.
\end {example}

\begin {example}
Let $X, Y$ be disjoint infinite countable sets,
  \begin {align*}
  E_0 &= \{A \subset (X \cup Y) \st \card (A \cap X) = \card (A \cap Y) \
           \text {is finite}\}\,,\\
  E   &= E_0 \cup \{(X \cup Y) \setminus A \st A \in E_0 \}\,,
  \end {align*}
$A \oplus B = A \cup B$ for disjoint $A,B \in E$. Then $(E, \oplus,
\emptyset, X \cup Y)$ is a concrete orthomodular poset with the maximality
property. Let $X = \{x_n \st n \in \N\}$, $y_0 \in Y$, $f \st X \to Y
\setminus \{y_0\}$ be a bijection, $A = (X \cup Y) \setminus \{x_1,
f(x_1)\}$, $B = (X \cup Y) \setminus \{x_1, y_0\}$. Then the chain $\bigl\{
\{ x_2, \dots, x_n, f(x_2), \dots, f(x_n) \} \st n \in \N \setminus \{1\}
\bigr\}$ in $[\emptyset, A] \cap [\emptyset, B]$ does not have an upper
bound in $[\emptyset, A] \cap [\emptyset, B]$, hence the condition~(CU) from
\ref {T:MaximalityProperties} is not fulfilled.
\end {example}

Let us remark that not all effect algebras have the maximality property (see
\ref {E:OMPnotM}).

\section *{Acknowledgements}

The work was supported by the grant of the Grant Agency of the Czech
Republic no.~201/07/1051 and by the research plan of the Ministry of
Education of the Czech Republic no.~6840770010.

\end {document}